# Mid-infrared pure-state quantum light source based on lithium niobate waveguides*

HUANG Yuhang[1], WANG Dongzhou[2], KE Shaolin[1, *], JIN Ruibo[1, 3, *]

1. Hubei Key Laboratory of Optical Information and Pattern Recognition, Wuhan Institute of Technology, Wuhan 430205, China
2. Jinan Institute of Quantum Technology, Jinan 250101, China
3. Key Laboratory of Low-Dimensional Quantum Structures and Quantum Control of Ministry of Education, Hunan Normal University, Changsha 410081, China

**Abstract** Mid-infrared quantum light sources hold broad application prospects in fields such as gas sensing and infrared thermal imaging. However, currently used mid-infrared quantum entangled light sources primarily rely on bulk periodically poled lithium niobate (PPLN) crystals, which limits brightness and integration. This paper proposes a theoretical scheme based on lithium niobate thin films, in which 1556.9 nm pumping is used to generate entangled photon pairs with a central wavelength of 3113.8 nm. By optimizing the waveguide structure and periodic polarization design, type-II phase matching and group velocity matching are achieved. This enables transverse electric (TE)-polarized pump input to be down converted to generate photon pairs with TE and transverse magnetic (TM) polarizations. Furthermore, by combining a domain arrangement algorithm used for the customized design of polarization direction in PPLN waveguides, precise phase matching is achieved, resulting in a quantum light source with a purity as high as 0.999 and a brightness of $6.18\times10^6$ cps/mW, which is three orders of magnitude higher than that of the bulk PPLN crystal source. This study provides a promising solution for realizing high-brightness, high-purity on-chip quantum light sources in the mid-infrared band.

---

* The paper is an English translated version of the original Chinese paper published in Acta Physica Sinica. Please cite the paper as: **HUANG Yuhang, WANG Dongzhou, KE Shaolin, JIN Ruibo, Mid-infrared pure-state quantum light source based on lithium niobate waveguides.** *Acta Phys. Sin.*, 2026, 75(4): 040602. DOI: 10.7498/aps.75.20251398

## 1 Introduction

Quantum light sources in the mid-infrared band (approximately 2-20 μm) play a pivotal role in quantum information science and technology. They have become key resources driving frontier applications such as quantum sensing, quantum communication, and quantum imaging[1-3]. This spectral range covers strong absorption lines of various gases, which significantly enhances the sensitivity and precision of gas sensing and environmental monitoring. Furthermore, the mid-infrared band includes the atmospheric transmission window between 3 and 5 μm. This window exhibits high transparency, providing low-loss channels for free-space quantum communication and secure quantum transmission[4,5]. In terms of imaging, objects at room temperature spontaneously emit mid-infrared radiation. Consequently, quantum sources in this band are suitable for infrared thermal imaging[6-8] and super-resolution, low-background imaging of biomedical samples[9,10]. These sources also serve high-demand scenarios such as military LiDAR[11] and chemical sample detection[12].

Currently, the primary method for generating mid-infrared single-photon sources is spontaneous parametric down-conversion (SPDC). In crystals with second-order nonlinear effects, high-frequency pump photons convert into low-frequency signal and idler photons when phase-matching conditions are met[13,14]. Lithium niobate possesses a large second-order nonlinear coefficient and exhibits excellent transparency from the visible to the mid-infrared range. Additionally, its mature electric field poling technology enables efficient quasi-phase matching. Therefore, periodically poled lithium niobate (PPLN) serves as an ideal material platform for realizing SPDC processes. Based on the polarization relationship between signal and idler photons, SPDC is classified into Type-0, Type-I, and Type-II. Each type holds significant value for various applications. For instance, Type-II SPDC naturally generates orthogonally polarized photon pairs, which can directly serve as polarization-entangled sources. Recent years have witnessed substantial progress in mid-infrared waveguide devices. In 2017, Mancinelli et al.[7] pumped a PPLN crystal with 1550 nm continuous-wave light. They generated degenerate quantum sources at 2890 nm and 3343 nm via SPDC. In 2020, Prabhakar et al.[15] experimentally excited a PPLN crystal using 1045 nm femtosecond pump pulses. They obtained polarization-entangled photon pairs at 2 μm. In 2021, Wei et al.[16] prepared spectrally uncorrelated biphotons in the mid-infrared band (MIR) using doped lithium niobate crystals through SPDC. In 2022, Zhang Chentao et al.[17] improved the purity of single photons generated in PPLN by designing lithium niobate crystals using domain arrangement algorithms. In 2023, Cai et al.[18] theoretically analyzed the feasibility of generating pure single photons in the mid-infrared range using 14 birefringent crystals and 8 periodically poled crystals. In 2024, Ge et al.[19] used Type-II phase-matched periodically poled potassium titanyl phosphate crystals. Under 1541 nm pumping, they generated degenerate photon pairs at 3082 nm. This work achieved the generation of time-energy entangled photon pairs in the mid-infrared band (3082 nm) for the first time. In 2025, Li et al.[20] developed a two-color polarization-entangled photon source at 3370 nm and 1555 nm. This development offers potential for all-weather operation in free-space quantum communication in the mid-infrared band. Zhu et al.[21] theoretically designed an ultra-broadband entangled biphoton source based on custom PPLN. They achieved a bandwidth of 4780 nm and attosecond-level Hong-Ou-Mandel (HOM) interference fringes. This provides critical light source support for next-generation quantum technology applications.

Previous studies utilized nonlinear bulk crystals to generate mid-infrared quantum sources. However, this approach suffers from low photon yield and poor integrability. To address these limitations, this paper designs a PPLN waveguide structure to replace bulk crystals. By satisfying group velocity matching conditions at 3.1 μm, we obtain high-brightness, high-purity quantum sources. Regarding applications, the compact size of waveguide devices facilitates integration with lasers, modulators, detectors, and other photonic components on a single chip[22,23]. This advantage is expected to promote the development of integrated quantum optical systems in the mid-infrared band[24-26].

# 2 Theory

In the SPDC process, one pump photon splits into two lower-energy photons: a signal photon and an idler photon. The biphoton component in SPDC can be expressed as[27]

$$|\psi_{\mathrm{si}}\rangle = \int_0^\infty \int_0^\infty \mathrm{d}\,\omega_{\mathrm{s}} \mathrm{d}\omega_{\mathrm{i}} f(\omega_{\mathrm{s}},\omega_{\mathrm{i}}) \hat{a}_{\mathrm{s}}^{\dagger}(\omega_{\mathrm{s}}) \hat{a}_{\mathrm{i}}^{\dagger}(\omega_{\mathrm{i}})|0\rangle, \quad (1)$$

where $f(\omega_{\mathrm{s}},\omega_{\mathrm{i}})$ denotes the joint spectral amplitude (JSA):

$$f(\omega_{\mathrm{s}},\omega_{\mathrm{i}}) = \phi(\omega_{\mathrm{s}},\omega_{\mathrm{i}})\alpha(\omega_{\mathrm{s}}+\omega_{\mathrm{i}}).$$

Here, $\phi(\omega_{\mathrm{s}},\omega_{\mathrm{i}})$ and $\alpha(\omega_{\mathrm{s}}+\omega_{\mathrm{i}})$ denote the phase-matching amplitude and the pump envelope amplitude, respectively. The periodic poling of the PPLN compensates for the phase mismatch $\Delta k$ induced by dispersion, as expressed by

$$\Delta k = 2\pi\left(\frac{n_{\mathrm{eff}}(\lambda_{\mathrm{p}})}{\lambda_{\mathrm{p}}} - \frac{n_{\mathrm{eff}}(\lambda_{\mathrm{s}})}{\lambda_{\mathrm{s}}} - \frac{n_{\mathrm{eff}}(\lambda_{\mathrm{i}})}{\lambda_{\mathrm{i}}} - \frac{1}{\Lambda}\right), \quad (2)$$

Here, $n_{\mathrm{eff}}(\lambda_{\mathrm{p(s,i)}})$ and $\lambda_{\mathrm{p(s,i)}}$ denote the modal effective refractive index and wavelength of the pump (signal, idler) light, respectively, while $\Lambda$ represents the poling period. According to the law of conservation of energy, $\omega_{\mathrm{s}}+\omega_{\mathrm{i}}=\omega_{\mathrm{p}}$. Consequently, the ridge of the pump envelope function $\alpha(\omega_{\mathrm{s}}+\omega_{\mathrm{i}})$ always aligns along the anti-diagonal direction. To ensure that the phase-matching function $\phi(\omega_\mathrm{s},\omega_\mathrm{i})$ distributes along the diagonal direction, the group velocity matching (GVM) condition must be satisfied[28]:

$$2V_{\mathrm{g,p}}^{-1}(\omega_\mathrm{p}) = V_{\mathrm{g,s}}^{-1}(\omega_\mathrm{s}) + V_{\mathrm{g,i}}^{-1}(\omega_\mathrm{i}), \qquad (3)$$

Here, $V_{\mathrm{p(s,i)}}^{-1}$ denotes the inverse of the group velocity for the pump (signal, idler) light. Under this condition, the joint spectral amplitude (JSA) exhibits a circular distribution. The group velocity is defined as

$$V_\mathrm{g} = \frac{\mathrm{d}\omega}{\mathrm{d}k} = \frac{c}{n_\mathrm{eff} - n_\mathrm{eff}'\lambda}, \qquad (4)$$

Here, $n_\mathrm{eff}'$ denotes the derivative of the effective refractive index $n_\mathrm{eff}$ with respect to the wavelength $\lambda$. The properties of the JSA can be measured using HOM interference. The expression for the coincidence probability $p$ as a function of the time delay $\tau$ is given as follows[29]:

$$p(\tau) = \frac{1}{2} - \frac{1}{2}\int_{-\infty}^{\infty}\int_{-\infty}^{\infty} |f(\omega_\mathrm{s},\omega_\mathrm{i})|^2 \cos(\omega_\mathrm{s} - \omega_\mathrm{i})\tau \mathrm{d}\omega_\mathrm{s}\mathrm{d}\omega_\mathrm{i}. \qquad (5)$$

Purity is a parameter describing the spectral correlation of a photon source. It can be calculated by performing a Schmidt decomposition on $f(\omega_\mathrm{s},\omega_\mathrm{i})$:

$$f(\omega_\mathrm{s},\omega_\mathrm{i}) = \sum_j c_j \phi_j(\omega_\mathrm{s})\varphi_j(\omega_\mathrm{i}), \qquad (6)$$

Here, $\phi_j(\omega_\mathrm{s})$ and $\varphi_j(\omega_\mathrm{i})$ represent two orthogonal basis vectors in the frequency domain. The term $c_j$ denotes a non-negative real number satisfying the normalization condition $\sum_j c_j^2 = 1$. The purity $\mathcal{P}$ is defined as

$$\mathcal{P} = \sum_j c_j^4. \qquad (7)$$

Due to the characteristics of the phase-matching function, side lobes exist in the joint spectral amplitude (JSA), which reduces the purity of the quantum light source. A design method involving customized poling periods can be employed to enhance purity, as detailed in Section 3.4.

# 3 Simulation

## 3.1 Design of PPLN Waveguides with Group Velocity Matching

Figure 1(a) illustrates the cross-section of the PPLN waveguide designed in this study, which features a ridge structure with light propagating along the $y$ direction. This waveguide is fabricated by etching a lithium niobate thin film on a silicon dioxide ($SiO_2$) buffer layer. The $h_2 = 2.3\mu$m layer, deposited on a silicon (Si) substrate, serves as an electrical insulator with a thickness of 10 μm, while the Si substrate provides mechanical support. The primary geometric parameters are defined as follows: $w_1$ denotes the ridge top width, $h_1$ represents the etching depth, $h_2$ indicates the residual thickness of the lithium niobate film after etching, and $\theta$ signifies the sidewall angle. Due to limitations in current micro-nano fabrication processes, achieving perfectly vertical sidewalls in lithium niobate is challenging; typically, the sidewall angle $\theta$ ranges from 40° to 80°. The geometric parameters adopted in this study are: $w_1 = 4070$nm, $h_1 = 1.5\mu$m, $\theta = 65°$, and $d_{15}$. Lithium niobate is an anisotropic crystal, and its refractive index is calculated using the Sellmeier equation reported by Gayer et al.[30]. This study employs z-cut lithium niobate crystals to fully utilize the $d_{15}$ component of the second-order nonlinear coefficient, thereby achieving Type-II phase matching (transverse electric (TE) polarization → TE polarization + transverse magnetic (TM) polarization).

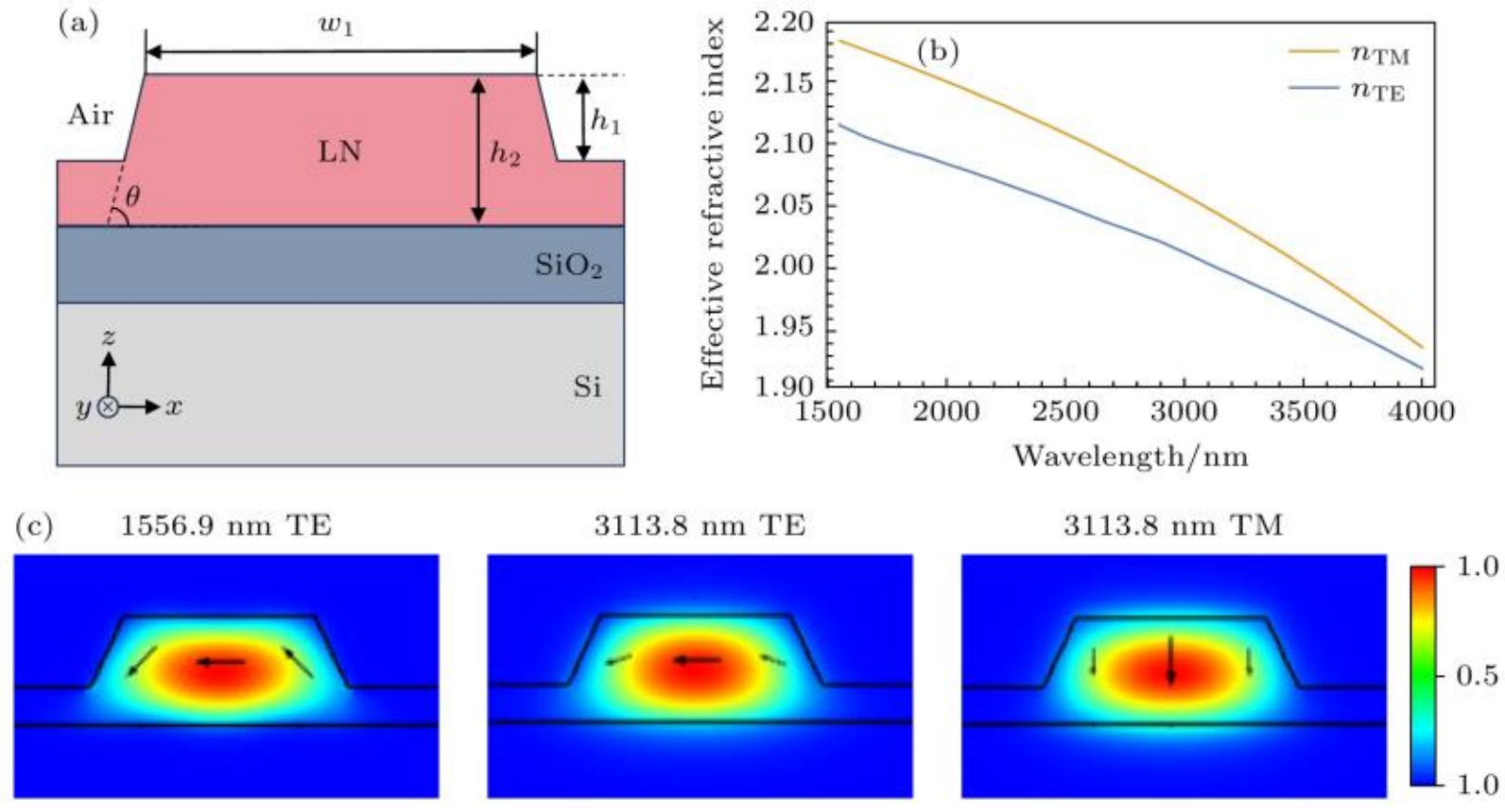

**Fig. 1 Waveguide mode profiles and dispersion: (a) Cross-section of the PPLN waveguide, where $w_1 = 4070$nm, $h_1 = 1.5\mu$m, $h_2 = 2.3\mu$m, $\theta = 65°$; (b) dispersion curves of TE and TM modes; (c) mode field profiles of pump and down-converted light, with arrows indicating polarization directions.**

To analyze the phase matching and group velocity matching conditions of the waveguide, this study employs the finite element method to numerically simulate the dispersion characteristics of TE and TM modes within the waveguide. The results are presented in Fig. 1(b). The calculation temperature is set to 50 ℃. Within the wavelength range of 1500—4000 nm, the effective refractive index of the TM mode exceeds that of the TE mode. Furthermore, the effective refractive indices of both modes decrease as the wavelength increases. Based on the calculated mode dispersion curves, the group velocity matching wavelength can be determined using Eq. (3). We consider the Type-II SPDC scenario, where a TE-mode pump photon undergoes down-conversion to generate a TE-mode signal photon and a TM-mode idler photon. Calculations indicate that the central wavelength for degenerate down-conversion satisfying the group velocity matching condition is 3113.8 nm, corresponding to a pump wavelength of 1556.9 nm. The poling period of the PPLN can be further determined based on the effective refractive indices at these wavelengths. Specifically, the effective refractive index of the TE-mode pump light at 1556.9 nm is $n_{\mathrm{eff}}(\mathrm{TE}, 1556.9\mathrm{nm}) = 2.1117$. The effective refractive index of the TE-mode signal light at 3113.8 nm is $n_{\mathrm{eff}}(\mathrm{TE}, 3113.8\mathrm{nm}) = 2.0027$, and that of the TM-mode idler light is $n_{\mathrm{eff}}(\mathrm{TM}, 3113.8\mathrm{nm}) = 2.0468$. Using Eq. (2), we obtain a poling period of $\Lambda = 17.9\mu\mathrm{m}$. Fig. 1(c) illustrates the mode distributions of the pump and down-converted light in the waveguide cross-section, with arrows indicating the polarization directions of the light.

### 3.2 Analysis of Spectral Distribution

To investigate the joint spectral characteristics of entangled photon pairs, we calculated the joint spectral amplitude (JSA) at the group velocity matching (GVM) wavelength of $\lambda = 3113.8$ nm. Fig. 2(a) depicts the pump envelope function (PEF). Its ridge is distributed along the anti-diagonal direction, and the full width at half maximum (FWHM) of the pump light is 0.58 nm. Fig. 2(b) shows the distribution of the phase matching function (PMF) within the same wavelength coordinate system. The ridge of the phase matching function appears as a straight line. The observed curvature arises from the waveguide structure, whereas the ridge would be straight for bulk crystals. At the center of the figure, its tilt direction is approximately orthogonal to that of the pump envelope function. The emerging fringe structure originates from the sinc function nature of the phase matching function. Multiplying Fig. 2(a) by Fig. 2(b) yields the JSA shown in Fig. 2(c). The JSA distribution exhibits a single main lobe with an approximately circular shape. This indicates that the generated photon pairs possess strong decoupling characteristics in the frequency domain. Numerical calculations yield a frequency-domain purity $\mathcal{P}$ of 0.82 for this JSA. This result suggests that the prepared two-photon state has good separability in the frequency degree of freedom. Furthermore, we evaluate the Hong-Ou-Mandel (HOM) interference visibility based on this JSA. Substituting the JSA into Eq. (5) produces the HOM interference fringes, as shown in Fig. 3(a2). The HOM interference fringes exhibit a triangular shape because the JSA behaves as a sinc function along the difference frequency direction. The measured HOM interference visibility is 1. This result reflects the excellent quantum coherence and spectral indistinguishability of the prepared quantum light source. It provides a crucial guarantee for high-fidelity operations in subsequent quantum information processing tasks.

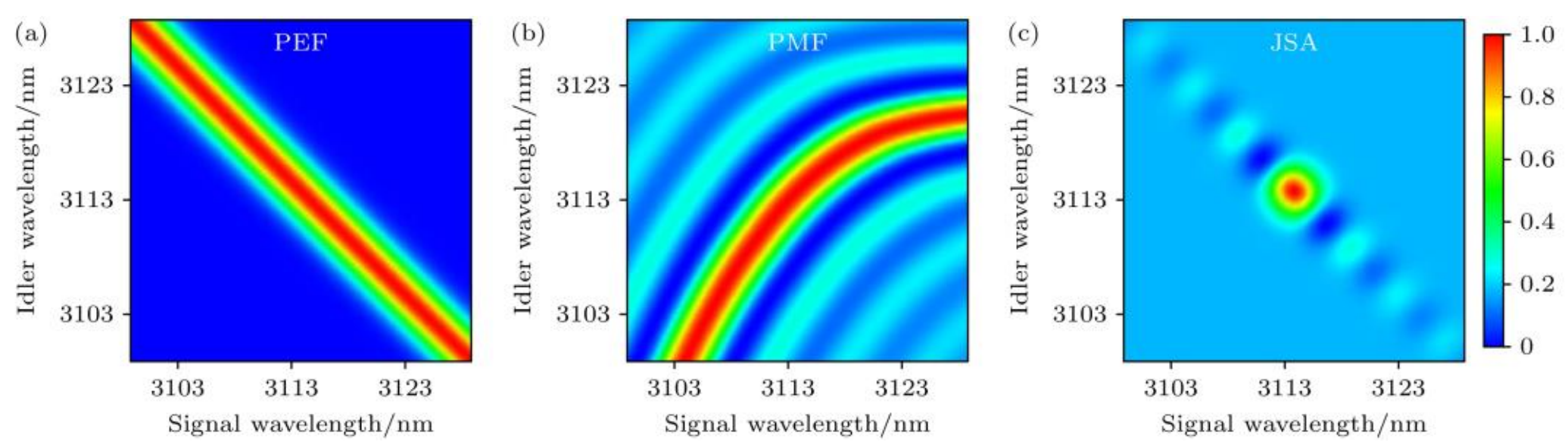

**Fig. 2 (a) Pump envelope function; (b) phase matching function; (c) joint spectral amplitude.**

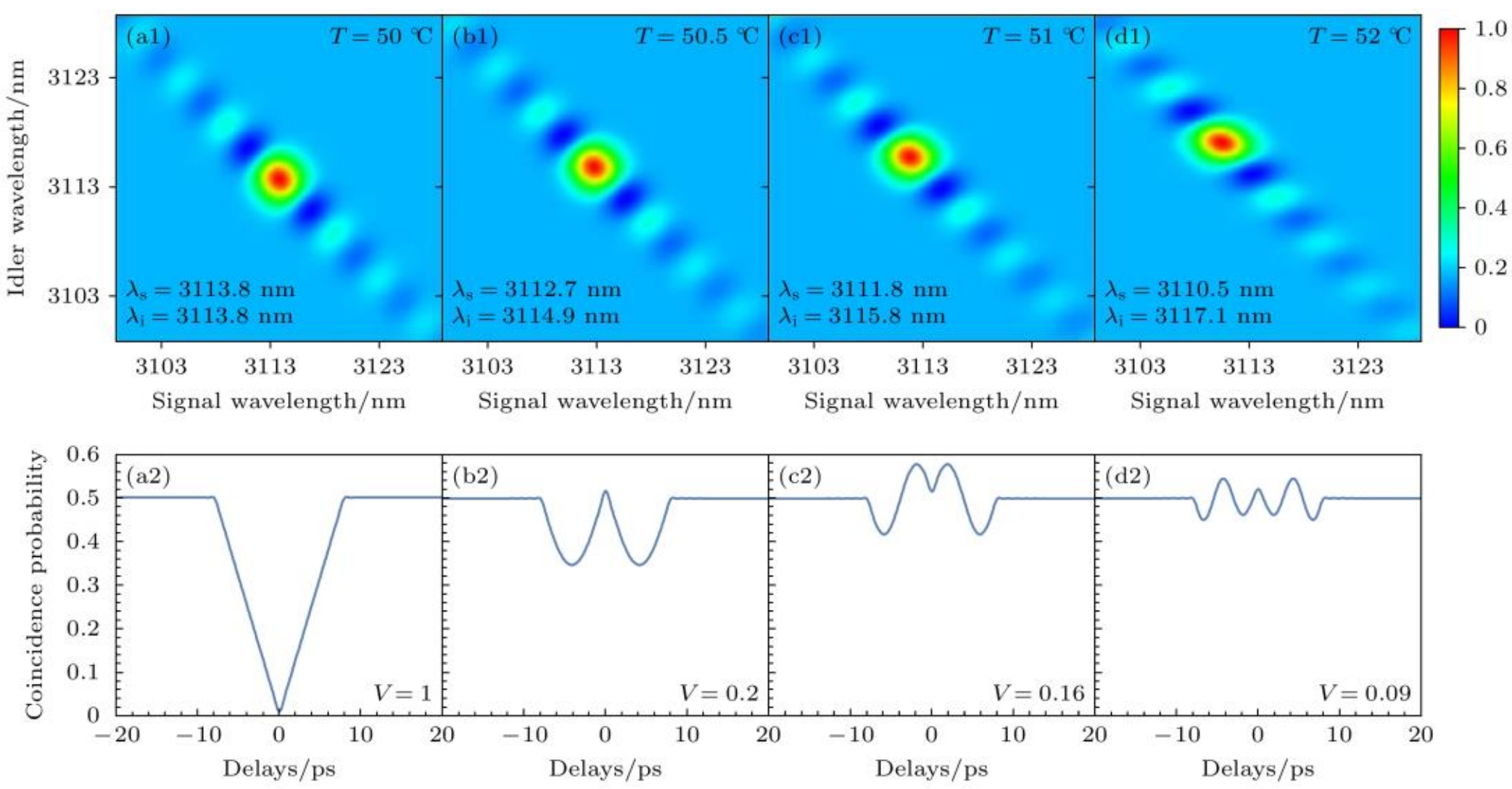

**Fig. 3 JSA (a1)-(d1) and HOM interference patterns (a2)-(d2) at different temperatures: (a) 50 ℃; (b) 50.5 ℃; (c) 51 ℃; (d) 52 ℃.**

## 3.3 Analysis of Temperature Effects

To investigate the environmental adaptability of the quantum light source, we systematically analyzed the impact of temperature variations on the two-photon spectral characteristics and Hong-Ou-Mandel (HOM) interference. Numerical simulations calculated the evolution of the joint spectral amplitude (JSA) and HOM interference visibility within the 50-52 ℃ temperature range, as shown in Figure 3. Figures 3(a1)-(d1) present the JSA at temperatures of 50 ℃, 50.5 ℃, 51 ℃, and 52 ℃, respectively, illustrating the JSA evolution as the temperature rises from 50 ℃ to 52 ℃. The thermo-optic effect causes temperature changes to alter the effective refractive index of the waveguide. This modification affects the phase-matching conditions, shifting the JSA toward the upper-left corner. The lower-left corner of each JSA plot displays the central wavelengths of the signal and idler photons. Figures 3(a2)-(d2) show the HOM interference curves corresponding to each temperature point. The HOM interference visibility decreases from 1 at 50 ℃ to 0.2, 0.16, and 0.09. The interference visibility is defined as $V = (p_{\max} - p_{\min})/(p_{\max} + p_{\min})$, where $p_{\max(\min)}$ represents the maximum (minimum) value of $p(\tau)$[31]. As the temperature increases by 2 ℃, the JSA deviates from a symmetric distribution, thereby reducing the HOM interference degree at the beam splitter. This indicates that this type of waveguide is highly sensitive to temperature. Consequently, precise temperature control devices are necessary for future practical applications.

According to the quantum version of the Wiener-Khinchin theorem[29], the HOM interference fringes and the projection of the two-photon joint spectral intensity (JSI) along the difference frequency direction ( $\omega_s - \omega_i$ ) exhibit a Fourier transform relationship. Here, JSI = $|f(\omega_s, \omega_i)|^2$. When the projection of the JSI in the difference frequency direction is a $\mathrm{sinc}^2$ function, its Fourier transform yields a triangular function. This corresponds to the scenario shown in Figures 3(a1) and (a2). When the JSI shifts along the difference frequency direction, transforming Figure 3(a1) into Figures 3(b1)-(d1), the Fourier transform property of translation applies. Consequently, the triangular function becomes modulated by a cosine term, which corresponds to the shapes observed in Figures 3(b2)-(d2).

### 3.4 Design of Customized Poling Periods

The purity of the quantum light source fabricated using the aforementioned waveguide structure at 50 °C is only 0.82. To further enhance this purity, we can customize the period of the lithium niobate waveguide. Using the domain arrangement algorithm described in references [17,32], we designed the periodic structure shown in Figure 4. The total length of the waveguide is 36 mm, with a domain width of $L_c = \Lambda/2$. Figure 4(a) details the poling distribution within the waveguide, with insets showing poling details along the 9.0-9.5 mm and 17.0-17.5 mm sections. By regulating the poling period structure, we suppress the side lobes in the waveguide phase-matching function. This suppression achieves frequency-uncorrelated two-photon states and facilitates the preparation of high-purity single-photon sources. The JSA in Figure 4(d) results from multiplying the pump envelope function in Figure 4(b) by the phase-matching function in Figure 4(c). Schmidt decomposition calculations yield a frequency-domain purity of 0.999 for the quantum light source.

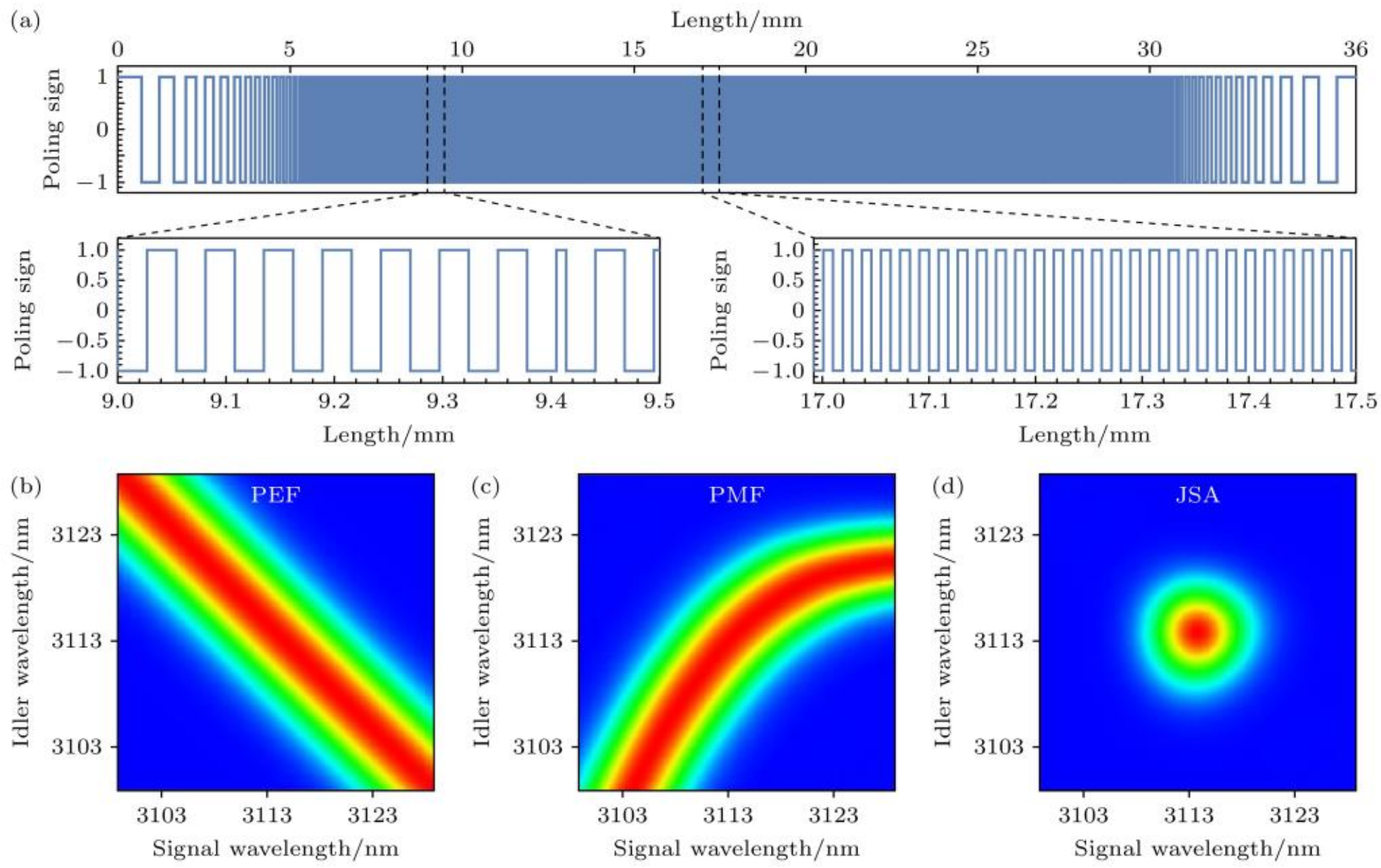

**Fig. 4 (a) Customized polarization distribution of the PPLN waveguide; (b) pump envelope function; (c) phase-matching function; (d) joint spectral distribution under the customized polarization period, achieving a quantum light source purity of 0.999.**

## 3.5 Estimation of Photon Pair Generation Rate

The photon pair generation rate serves as a key metric for evaluating the performance of quantum light sources. This section focuses on calculating this parameter. We first determine the effective mode area of the waveguide, which equals the reciprocal of the nonlinear coupling coefficient. The expression for the nonlinear coupling coefficient $\Gamma$ is given as[33]

$$\Gamma = \frac{|\int E_{\mathrm{p}} E_{\mathrm{s}}^{*} E_{\mathrm{i}}^{*} \mathrm{d}x\mathrm{d}z|^{2}}{\int |E_{\mathrm{p}}|^{2}\mathrm{d}x\mathrm{d}z \int |E_{\mathrm{s}}|^{2}\mathrm{d}x\mathrm{d}z \int |E_{\mathrm{i}}|^{2}\mathrm{d}x\mathrm{d}z}. \tag{8}$$

Equation (8) calculates the overlap integral among three modes: the $x$-component of the electric field for the pump light TE mode, the $x$-component for the down-converted light TE mode, and the $z$-component for the TM mode. The calculation yields a nonlinear coupling coefficient of $\Gamma = 0.1316\mu\mathrm{m}^{-2}$, with the effective mode area in the waveguide defined as $A_{\mathrm{eff}} = 1/\Gamma$. Subsequently, we estimate the photon pair generation rate. Under type-II SPDC conditions, the photon pair generation rate is given by[34,35]

$$\begin{aligned} R_{\mathrm{sm}} &= \frac{P}{4\varepsilon_0 \pi c^3} \frac{n_{\mathrm{gs}} n_{\mathrm{gi}}}{n_{\mathrm{eff}}^2(\lambda_{\mathrm{s}}) n_{\mathrm{eff}}^2(\lambda_{\mathrm{i}}) n_{\mathrm{eff}}(\lambda_{\mathrm{p}})} d_{\mathrm{eff}}^2 \frac{1}{A_{\mathrm{eff}}} \\ &\times \int_0^{\omega_{\mathrm{p}}} \omega_{\mathrm{s}}(\omega_{\mathrm{p}} - \omega_{\mathrm{s}}) |\int_0^{L} g(z) \mathrm{e}^{-\mathrm{i}kz} \mathrm{d}z|^2 \mathrm{d}\omega_{\mathrm{s}}, \end{aligned} \tag{9}$$

Here, $\varepsilon_0$ denotes the vacuum permittivity, $n_{\mathrm{gs}}(n_{\mathrm{gi}})$ represents the group refractive index of the signal (idler) light, and $d_{\mathrm{eff}}$ is the effective nonlinear coefficient, where $d_{\mathrm{eff}} = d_{15} = -4.6\mathrm{pm/V}$. Additionally, $\omega_{\mathrm{p}}(\omega_{\mathrm{s}})$ indicates the frequency of the pump (signal) light, and $g(z)$ defines the polarization direction. Calculations yield a photon pair generation rate of 6.18×10$^6$ cps at a pump power of 1mW.

Next, we compare this with bulk PPLN crystals. The expression for the effective mode field area of bulk PPLN crystals is given by

$$A_{\mathrm{eff}} = \frac{\pi}{8}\left|\frac{\sigma_1^2+2\sigma_{\mathrm{p}}^2}{\sigma_{\mathrm{p}}}\right|^2, \qquad (10)$$

Here, $\sigma_{\mathrm{p}(1)}$ denotes the beam waist radius of the pump light (down-converted light). For a bulk PPLN crystal with a common diameter of 100 μm, the effective mode field area is approximately $8835.73\mu\mathrm{m}^2$. In contrast, the effective mode field area for the waveguide structure designed in this study is approximately $7.599\mu\mathrm{m}^2$. This represents a reduction to roughly 1/1000 of the bulk crystal value, corresponding to a three-order-of-magnitude increase in the photon pair generation rate[35].

# 4 Conclusion

This study achieved Type-II phase matching and group velocity matching in the 3.1 μm band through the design of the waveguide structure and poling period. This approach significantly enhanced both the photon pair generation rate and spectral purity. Numerical simulations indicate that the designed PPLN waveguide efficiently generates TE/TM polarization-entangled photon pairs at 3113.8 nm under TE-mode pumping at 1556.9 nm. The photon pair generation rate increases by three orders of magnitude compared to traditional bulk PPLN crystals, reaching 6.18×10$^6$ cps/mW. In Hong-Ou-Mandel (HOM) interference, the visibility can reach unity. Furthermore, temperature tuning allows for the modulation of the joint spectral amplitude (JSA) and HOM interference fringes. We further employed a domain engineering algorithm to customize the poling period. This design effectively suppressed the sidelobes of the phase-matching function, resulting in a highly symmetric circular distribution of the joint spectral amplitude and improving the spectral purity to 0.999. The pure-state quantum light source generated in this work offers advantages in integrability and miniaturization. It holds significant potential for future applications in quantum sensing, quantum imaging, and quantum communication[36,37].